# Cation vacancy order in the $K_{0.8+x}Fe_{1.6-y}Se_2$ system: five-fold cell expansion accommodates 20% tetrahedral vacancies

J. Bacsa[a], A.Y. Ganin[a], Y. Takabayashi[b], K.E. Christensen[c], K. Prassides[b], M.J. Rosseinsky[a], J.B. Claridge[a],*

Ordering of the tetrahedral site vacancies in two crystals of refined compositions $K_{0.93(1)}Fe_{1.52(1)}Se_2$ and $K_{0.862(3)}Fe_{1.563(4)}Se_2$ produces a fivefold expansion of the parent $ThCr_2Si_2$ unit cell in the ab plane which can accommodate 20% vacancies on a single site within the square FeSe layer. The iron charge state is maintained close to +2 by coupling of the level of alkali metal and iron vacancies, producing a potential
10 doping mechanism which can operate at both average and local structure levels.

## Introduction

The mechanisms of charge carrier density control in the iron pnictide and chalcogenide superconductors[1,2] are important as small changes in composition produce metal-insulator transitions
15 and generate superconductivity at temperatures of up to 37K in chalcogenides[3,4] and 55K in pnictides[5,6]. All of the reported materials are based on a square FeX (X = Se, As) layer built from edge-sharing of $FeX_4$ tetrahedra. The negative charge on the layers in the pnictide case requires the presence of charge
20 compensating cations (LiFeAs) or cation-oxide slabs (LaOFeAs). The recent demonstration that insertion of alkali metal cations between FeSe layers affords superconductivity in $A_xFe_{2-y}Se_2$ (A = K, Rb, Cs, Tl: reported compositions for $0.7 < x < 1$, $0 < y < 0.5$)[7-9] materials places the defect chemistry in sharp relief, as
25 attaining an iron charge state close to +2 found in the other Fe-based superconducting families will require the creation of considerable defect concentrations on both iron and alkali metal sites. Such densities suggest defect assimilation or elimination mechanisms will be required with the originally stoichiometric
30 FeSe layers, and these may be reflected locally in other related superconducting families where lower defect densities are needed to tune electronic properties. The crystal structures of the $A_xFe_{2-y}Se_2$ (A = K, Rb, Cs) systems have been invariably described as the tetragonal $ThCr_2Si_2$ structure (space group $I4/mmm$, with the
35 same square FeSe lattice as in FeSe itself where $a = b \approx 3.9$ Å) with a random distribution of alkali metal and/or iron defects. Vacancy ordering of the Fe cations was proposed to lead to a 5 × 5 × 1 superstructure[10], with a range of modulation vectors observed by TEM and associated with defect ordering.[11] Here we
40 present the crystal structure of two closely compositionally related members of this family as determined by single crystal X-ray diffraction.

## Results and Discussion

In an Ar-filled glove box, 15.4 mg (0.394 mmol) of K metal
45 (Aldrich, 99.95%) was mixed in an alumina crucible with 106.2 mg (0.788 mmol) of FeSe, which was synthesized following the procedure described in [12]. The crucible was loaded into a 12-mm silica ampoule, which was sealed under reduced Ar atmosphere

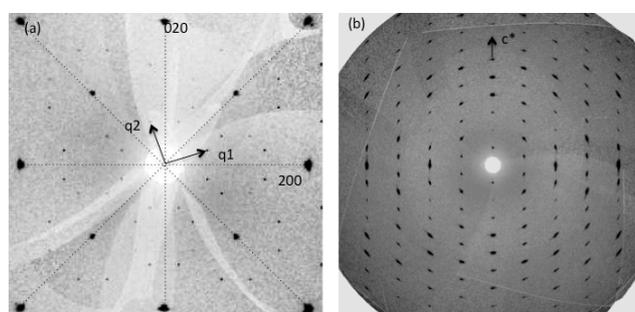

50 **Figure 1**: Reconstructed precession photographs for crystal 2 in the (a) *hk0* and (b) *h0l* layers. (a) shows the two modulation vectors required to describe the cell and the sub-cell twin planes that relate the two twin components are shown as a dotted lines.

55 (ca. 500 Torr). The ampoule was heated from ambient temperature to 1030°C at 1°C/min, and after 2 h of reaction time was cooled at 4°C/min to 700°C. The furnace was then switched off and the ampoule was allowed to cool naturally to ambient temperature. No visual evidence for ageing of the silica ampoule
60 due to reaction with the alkali metal could be observed. The ampoule was opened in the glove box to reveal a product consisting of fine shiny platelets. Magnetic measurements with a SQUID magnetometer at an applied field of 20 G on the bulk samples (3.4 mg) and single crystals (mounted using silicone
65 grease on a PTFE plate on a Teflon rod) from this batch showed they were non-superconducting. The crystals were not exposed to air at any point.

Single crystal diffraction data were collected at 100K on a Rigaku MicroMax-007HF rotating anode source (crystal 1) with Mo
70 radiation and at 90K on beamline I19 (crystal 2) at the Diamond Light Source using the Rigaku CrystalLogic Kappa goniometer at the zirconium absorption edge ($\lambda$= 0.6889 Å). Both diffractometers used the Saturn 724+ detector. Crystals were transferred into either Fomblin-Y or Paratone-N oil and cut to
75 size prior to mounting on MiTeGen loops. At I19, screening images on crystal 2 (0.05 x 0.05 x 0.03 $mm^3$) were collected with a scan width of 1 degree 5 degrees apart. The rectangular crystals had two sides of almost equal length and were smaller in the third direction. The crystals displayed a tetragonal diffraction pattern
80 $ThCr_2Si_2$ parent[10]. Detailed examination of reconstructed precession images of the *hk0* layer (Figure 1a) shows strong diffraction maxima due to the small sub-cell and additional weak peaks forming distinctive groups of eight reflections around the allowed peaks. The modulation is commensurate with $q_1 = (3/5a^* + 1/5b^*)$ and $q_2 = (-1/5a^* + 3/5b^*)$[11], where $a^*$ and $b^*$ refer to the *I4/mmm* subcell. The superstructure is commensurate, though as will be discussed later it does not have to be, depending on the precise iron vacancy level. The observed diffraction pattern can be accounted for as a twofold twinning about the original mirror planes perpendicular to the four fold axis of the sub-cell (dotted lines in Figure 1(a)) of a body centered **a**+2**b**, -2**a**+**b**, **c** cell which is a √5 × √5 × 1 expansion of the $ThCr_2Si_2$ cell. Elongation of reflections is observed along the interlayer direction **c**\* (Figure 1(b)). Based on the single crystal diffraction alone it is also possible to index the diffraction pattern in terms an untwined structure with 3 modulation vectors with $q_1 = (3/5a^* + 1/5b^*)$, $q_2 = (-1/5a^* + 3/5b^*)$ and $q_3 = (1/5a^* + 3/5b^*)$ or a single modulation vector $q_1 = (3/5a^* + 1/5b^*)$ implying monoclinic symmetry and four twin domains. The former, which would produce a 5 x 5 x 1 supercell in the commensurate approximation, was discounted due to the absence of second order $q_1+q_3$, $q_2+q_3$ modulation peaks and the observation of single domain electron diffraction patterns requiring fewer than three modulation vectors by Wang *et al.*[11]

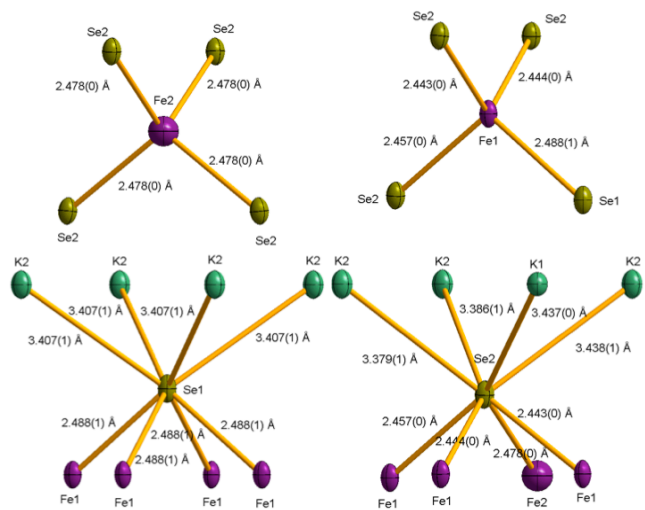

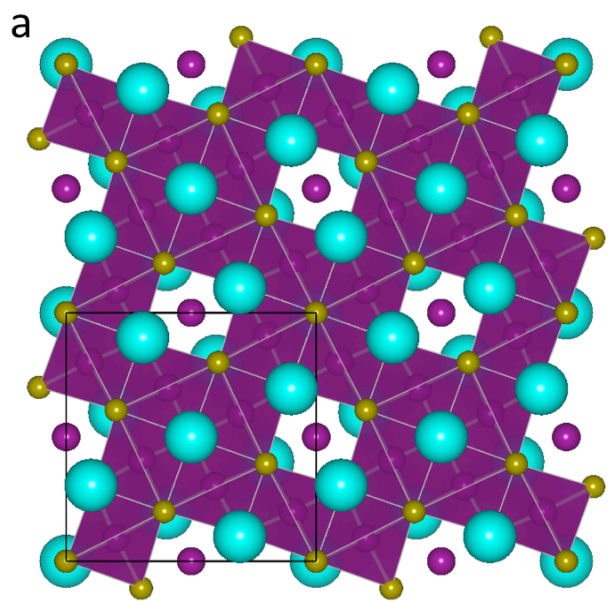

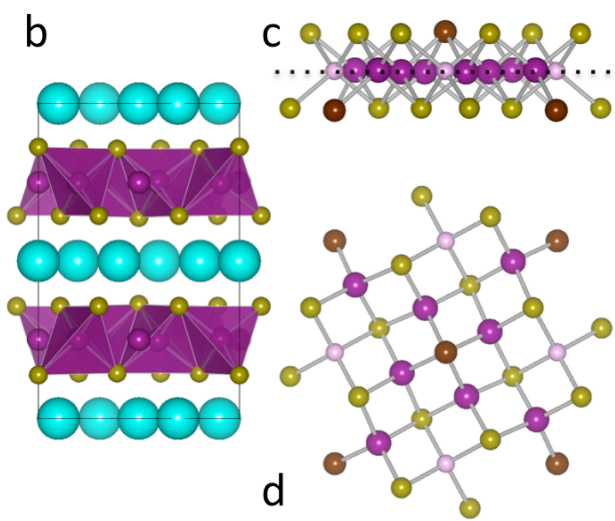

**Figure 2:** The crystal structure of $K_{0.862(3)}Fe_{1.563(4)}Se_2$ (a) View along [001] of the vacancy-containing $Fe_{1.6-x}Se_2$ plane. Fe1 sites (92.0(2)% occupied) represented as tetrahedral, Fe2 vacancy dominated sites as purple spheres. Se gold, K green spheres (b) K location between $Fe_{1.6-x}Se_2$ layers viewed perpendicular to c. (c) View along [010] showing a single $Fe_{1.6-x}Se_2$ layer. Tetrahedral Fe2 are pale pink and the four coordinate Se1 as brown sphere; the dotted line is drawn at z = ¼. (d) View along [001] of the same layer.

**Figure 3:** Coordination environments of the Fe and Se sites in $K_{0.862(3)}Fe_{1.563(4)}Se_2$ Fe sites are shown as purple, Se gold, and K green ellipsoids. Ellipsoids are drawn at 50%

Systematic absences of the individual domains are consistent with space group *I4/m* or *I112/m*. The data for crystal 2 were integrated on the $5a × 5b × c$ cell which indexes all the peaks for the twin domains and an empirical absorption correction based on redundancy was performed by abspack within the CrysAlis software suite. Attempts to treat the data as a twin directly in either a supercell or modulated description were either not possible or yielded unphysical results, in the limit where the two/four domains are due to micro twinning the results would be identical. The compositions obtained below using this absorption correction also are with in the composition range observed in a Jeol 2000FX TEM by EDX suggesting together with the prevalence of twinned electron diffraction patterns observed that this is a valid assumption. This reflection file was then decomposed using Jana2006[13] into a final reflection file for the two twin domains. Data on crystal 1 (0.12 x 0.13 x 0.05 $mm^3$) from the same batch were collected on the Rigaku rotating anode source and processed in an analogous manner using an analytical face-indexing absorption correction. Both crystal structures were solved using SUPERFLIP[14] and subsequently refined using Jana2006, solutions in *I112/m* were indistinguishable from those in *I4/m* and only the later will be discussed. Crystal 2 (I19 diffractometer) refined to give $R_1(obs)$ = 5.7%, $R_w(all)$ = 9.4% whilst crystal 1 (Rigaku diffractometer) refined to $R_1(obs)$ = 5.8%, $R_w(all)$ = 8.5%. The refined compositions were $K_{0.93(1)}Fe_{1.52(1)}Se_2$ (crystal 1) and $K_{0.862(3)}Fe_{1.563(4)}Se_2$ (crystal 2) corresponding to formal iron charge states of 2.02(2) and 2.008(7), respectively (Table 1). The different K and Fe vacancy levels in the two crystals thus give rise to the same iron charge state of +2 within experimental error. Bond lengths and angles (Table 2) derived from the I19 synchrotron X-ray data on crystal 2 are discussed further here.

The refined structure is shown in a polyhedral representation in Figure 2. The materials adopt a tetrahedral vacancy-ordered derivative of the ThCr$_2$Si$_2$ structure. The observed tetragonal cell is a five-fold expansion in the *ab* plane of the unit cell of the simple square layer found in FeSe[15] and other Fe-based superconductors (Figure 2), with a low occupancy site (the 11.8(7)% (crystal 1)/22.7(3)% (crystal 2) occupied Fe2 site) surrounded by a square of four nearly fully occupied sites (91.8(5)% (crystal 1)/92.0(1)% (crystal 2) occupied Fe1 site). This cell expansion is due to ordering of the high concentration of defects within the layer to minimise the associated strain. Considering the limiting case where the Fe2 site is empty and Fe1 fully occupied gives a maximum iron content for this vacancy distribution of Fe$_{1.6}$Se$_2$. The observed ordering is thus consistent with the refined compositions of both crystals as it corresponds to accommodating higher iron contents than the motifs reported at the widely observed M$_{1.5}$Se$_2$ composition for this layer in chalcogenides[16, 17] and oxychalcogenides[18]. It should be noted that early studies of the TlFe$_x$S$_2$ system suggested cells of this type without refinement of a model.[19] In K$_{0.862(3)}$Fe$_{1.563(4)}$Se$_2$ each Fe1 site has three occupied Fe1 sites (two at 2.7089 (5)Å and one at 2.9041 (5) Å) and one vacancy Fe2 site (at 2.7709 (4) Å) as neighbours. The Fe-Fe separation in the square layer in FeSe is 2.661Å at 100K, indicating that K intercalation coupled with the presence of vacant tetrahedral sites expands the layer. The vacancy distribution in each FeSe layer is identical producing channels of vacancies parallel to *c*.

The Fe1 site has *1* symmetry and is tetrahedrally coordinated by three Se1 and one Se2 anions (Figure 3), with a mean distance of 2.45(2)Å (the standard deviation of the distribution of distances, rather than the esd, is given following in brackets here). The low point symmetry of *1* at the nearly fully occupied Fe1 site will fully lift the degeneracy of the *3d* orbitals. The vacancy-dominated Fe2 site has $\bar{4}$ symmetry with four equivalent Fe-Se2 contacts of 2.477 Å (Figure 3), reflecting expansion of the average structure of the layer locally at the vacancies. There is a single Fe-Se distance of 2.382Å in the $\bar{4}$ symmetry FeSe$_4$ tetrahedra in FeSe at 100K, again suggesting K insertion places the layers under tension. There are two distinct anion sites in the layer (Figure 2). Se1 is located on a fourfold axis and makes four equivalent contacts to Fe1 in the layer, and four longer contacts to

**Table 1** Refined structural parameters of K$_{0.862(3)}$Fe$_{1.563(4)}$Se$_2$ at 90 K

| Atom | Site | Symmetry | x | y | z | U$_{eq}$ / Å$^2$ | Occupancy |
|------|------|----------|---|---|---|------------------|-----------|
| K1 | 2b | *4/m..* | 0.5 | 0.5 | 0 | 0.0159(4) | 0.855(5) |
| K2 | 8h | *m..* | 0.30560(9) | 0.90051(9) | 0 | 0.0198(3) | 0.864(3) |
| Fe1 | 16i | *1* | 0.30216(4) | 0.40718(4) | 0.24756(3) | 0.01612(11) | 0.9201(18) |
| Fe2 | 4d | *-4..* | 0.5 | 0 | 0.25 | 0.0323(12) | 0.227(3) |
| Se1 | 4e | *4..* | 0 | 0 | 0.13810(4) | 0.01709(11) | 1.00 |
| Se2 | 16i | *1* | 0.39173(3) | 0.20011(3) | 0.144098(18) | 0.01743(8) | 1.00 |

Data from the I19 diffractometer at Diamond Light Source refined on crystal 2. Space group number 87 *I4/m* a = 8.7653(2)Å c = 13.8811(5)Å V = 1066.49(5)Å$^3$ λ = 0.6889Å θ = 1.7–32.1° μ = 21.03 mm-1 *T* = 90 K 21490 reflections 36 parameters 5005 reflections with *I* > 3σ(*I*) *R*[*F*$^2$ > 2σ(*F*$^2$)] = 0.057 *wR*(*F*$^2$) = 0.094 Weighting scheme based on measured s.u.'s *w* = 1/(σ$^2$(*F*) + 0.0001*F*$^2$) *S* = 1.13 A search of Crystal Web for I4/m with 10% tolerance on the cell parameters gave no isostructural materials - Tl$_5$Te$_3$ and an antimony analogue have similar metrics but a different structure. This structural motif was proposed but not refined by Berger and co-workers[19] based on powder superstructures in the TlFeS system.

**Table 2** Bond lengths (Å) and angles (°) in K$_{0.862(3)}$Fe$_{1.563(4)}$Se$_2$ at 90 K: angles in ***bold italic*** are bisected by the *c* axis rather than the *ab* plane

| Atoms | Distance | Atoms | Angle |
|-------|----------|-------|-------|
| Fe1-Se1 | 2.4876(5) | Se1-Fe1-Se2 | 113.195(15) |
| Fe1-Se2 | 2.4440(4) | Se1-Fe1-Se2 | 113.224(15) |
| Fe1-Se2 | 2.4432(4) | ***Se1-Fe1-Se2*** | ***102.524(17)*** |
| Fe1-Se2 | 2.4566(4) | ***Se2-Fe1-Se2*** | ***107.938(16)*** |
|  |  | Se2-Fe1-Se2 | 107.318(15) |
|  |  | Se2-Fe1-Se2 | 112.504(15) |
| Fe2-Se2 | 2.4776(3) × 4 | ***Se2-Fe2-Se2*** | ***107.211(8) × 2*** |
|  |  | Se2-Fe2-Se2 | 110.613(8) × 6 |
| K1-Se2 | 3.4367(3) × 8 | Se2-K1-Se2 | 108.815(6) × 4 |
|  |  | Se2-K1-Se2 | 70.199(6) × 8 |
|  |  | Se2-K1-Se2 | 180.0(5) × 4 |
|  |  | Se2-K1-Se2 | 71.185(6) × 4 |
|  |  | Se2-K1-Se2 | 109.801(6) × 8 |
| K2-Se1 | 3.4074(7) × 2 |  |  |
| K2-Se2 | 3.3864(7) × 2 |  |  |
| K2-Se2 | 3.4376(7) × 2 |  |  |
| K2-Se2 | 3.3790(7) × 2 |  |  |

The 17 inequivalent bond angles at the K2 site are given in the CIF – they represent slight distortions from the angles found for K1 due to the lower site symmetry and are not quoted here.

a 90° rotated square of K2 cations in the interlayer space (Figure 3). Se2 neighbours an Fe2 site and makes three contacts to the highly occupied Fe1 sites, and four longer contacts to a square consisting of three K2 and one K1 cations in the interlayer space. Se2 bridges the edges of two pairs of Fe1Se4 tetrahedra and corner links two more distant Fe1 neighbours, while Se1 bridges four shared tetrahedral edges.

The potassium cations occupy eight coordinate sites within the AAA stacked $Fe_{1.6-y}Se_2$ layers, and are in cubic coordination similar to that in the fluorite structure. There are two distinct sites – K1 on the 2b position with $I4/m$ symmetry is coordinated solely by Se2 and sits between squares of four occupied Fe1 sites. K2 is coordinated by six Se2 and two Se1 anions, the Se1 sites being directly above each other along the $c$ direction, and is located above three Fe1 and one almost vacant Fe2 site. Unlike the two Fe sites, both K sites have very similar occupancies in both refined crystals, consistent with the longer bond lengths at the sites producing less of an energy penalty for vacancy disorder.

The distinction between the two Se sites rumples the surface of the $Fe_{1-x}Se$ layer giving two distinct thicknesses for the chalcogenide layers - a parameter shown to strongly influence $T_c$ in the vacancy-free $Fe_{1+x}(Se,Te)$ systems[20, 21] - of 3.10Å (Se1 to Se1 vertical separation along $c$), 3.02 Å (Se2 to Se1) and 2.94Å (Se2 to Se2) due to the greater displacement of Se2 from the predominantly vacancy bearing Fe2 site it neighbours. The Se2…Se2 distances of 3.989 Å parallel and 4.074 Å perpendicular to the $Fe_{1-x}Se$ plane reflect a strong compression of the Fe2Se4 tetrahedra within the basal $ab$ plane, also seen in the Fe1Se4 units. The Fe1-based tetrahedron has a mean Se-Fe-Se angle of 109(4)° - the distortion is more complex than found in simpler stoichiometric layers as although the two angles bisected by the $c$ axis are on average more compressed (105(3)°) than those bisected by the ab plane (111(2)°) as found in the simpler stoichiometric materials such as FeSe[15] and LiFeAs[22], the $ab$-bisected Se2-Fe1-Se2 angle of 107.32(2)° is smaller than that bisected by $c$ of 107.94(2)°, which is the typical case in the LnOFeAs systems[23]. The slab thicknesses discussed above are comparable to those for the undistorted FeSe. The introduction of K into the fluorite-like slabs increases the interlayer separation to between 3.83Å (Se1…Se1 contacts) and 4.00Å (Se2…Se2) from 2.55 Å, which should increase the electronic two-dimensionality of the system. However, as the packing motif of the FeSe layers along c in the two structural types is different, the interlayer Se...Se contacts are not affected much (3.75 Å in FeSe).

The refined fractional occupancies of both the Fe1 and Fe2 sites which are respectively full and empty in the ideal $Fe_{1.6}Se_2$ structure corresponding to the observed unit cell suggest that the crystals contain disordered regions. This may be due to the presence of fully disordered regions (11% occupancy of Fe2 corresponds to 15% of a disordered component with 80% occupancy of both sites in crystal 1) or to stacking faults where each layer contains the same vacancy ordering motif but phase coherence is lost in the stacking along $c$. This is consistent with the observation of extension of the reflections along $c^*$ in the observed diffraction patterns (Figure 1(b)) and suggests an order-disorder transition which could be controlled by annealing or subsolidus routes to these materials.

## Conclusions

The presence of two distinct iron sites in the $Fe_{1.6-y}Se_2$ layers suggests selective metal doping will be possible, with mixed anion materials accessible due to the two distinct coordination numbers of the two observed anion sites.

The crystal studied here are not superconducting but the role of vacancies in superconducting compositions is clear, and the present structures show how these vacancies can be accommodated. The ordered distribution of vacancies on the Fe sites is required to accommodate the high level of Fe deficiency compared with the FeSe parent material and other families of Fe-based superconductors. The structural motif observed is suited to a 4:1 occupied:vacancy ratio with an ideal layer composition of $Fe_{1.6}Se_2$. The observed compositions can be thought of as corresponding to $K_{0.8+x}Fe_{1.6-x/2}S_2$ to maintain the $Fe^{2+}$ charge state, with deviations from this relationship permitting electron or hole doping. Higher iron contents than 1.6 can be expected to produce alternative, sparser vacancy orderings based on different modulation vectors[11]. It is straightforward to envisage a family of materials with related tetrahedral site vacancy orderings allowing tuning of the metal composition within an ordered structure in turn giving different accessible electronic properties. The vacancies on the K sites are in contrast positionally disordered suggesting lower temperature synthesis approaches may allow enhanced control of their distribution with associated impact on the electronic behaviour. The observed compositions and disorder of both refined crystals do not completely fit with the underlying lattices suggesting milder synthesis conditions to obtain the same target compositions may afford better ordered materials.

## Acknowledgements

We thank EPSRC for support under EP/H000925/1, EP/G037132 and EP/G037949 and STFC for access to the synchrotron X-ray facilities at Diamond Light Source.

## Notes and references

[a] *Department of Chemistry, University of Liverpool, Liverpool L69 7ZD, UK E-mail:claridge@liv.ac.uk*
[b] *Department of Chemistry, Durham University, Durham DH1 3LE, UK*
[c] *Diamond Light Source, Harwell Science and Innovation Campus, OX11 0DE UK*

† Electronic Supplementary Information (ESI) available: CIF for crystals 1 and 2. See DOI: 10.1039/b000000x/

1. Y. Kamihara, T. Watanabe, M. Hirano and H. Hosono, *J. Am. Chem. Soc.*, 2008, **130**, 3296-+.
2. F. C. Hsu, J. Y. Luo, K. W. Yeh, T. K. Chen, T. W. Huang, P. M. Wu, Y. C. Lee, Y. L. Huang, Y. Y. Chu, D. C. Yan and M. K. Wu, *Proc. Nat. Acad. of Sci.*, 2008, **105**, 14262-14264.
3. S. Medvedev, T. M. McQueen, I. A. Troyan, T. Palasyuk, M. I. Eremets, R. J. Cava, S. Naghavi, F. Casper, V. Ksenofontov, G. Wortmann and C. Felser, *Nature Mate.*, 2009, **8**, 630-633.
4. S. Margadonna, Y. Takabayashi, Y. Ohishi, Y. Mizuguchi, Y. Takano, T. Kagayama, T. Nakagawa, M. Takata and K. Prassides, *Phys. Rev. B*, 2009, **80**, 064506.
5. X. H. Chen, T. Wu, G. Wu, R. H. Liu, H. Chen and D. F. Fang, *Nature*, 2008, **453**, 761-762.
6. Z. A. Ren, W. Lu, J. Yang, W. Yi, X. L. Shen, Z. C. Li, G. C. Che, X. L. Dong, L. L. Sun, F. Zhou and Z. X. Zhao, *Chinese Phys. Lett.*, 2008, **25**, 2215-2216.
7. J. Guo, S. Jin, G. Wang, S. Wang, K. Zhu, T. Zhou, M. He and X. Chen, *Phys. Rev. B*, 2010, **82**, 180520.
8. A. Krzton-Maziopa, Z. Shermadini, E. Pomjakushina, V. Pomjakushin, M. Bendele, A. Amato, R. Khasanov, H. Luetkens and K. Conder, *arXiv*, 2011, 1012.3637.
9. Y. Mizuguchi, H. Takeya, Y. Kawasaki, T. Ozaki, S. Tsuda, T. Yamaguchi and Y. Takano, *Appl. Phys. Lett.*, 2011, **98**, 042511.
10. P. Zavalij, W. Bao, X. F. Wang, J. J. Ying, X. H. Chen, D. M. Wang, J. B. He, X. Q. Wang, G. F. Chen, P.-Y. Hsieh, Q. Huang and M. A. Green, *arXiv:*, 2011, 1101.4882.
11. Z. Wang, Y. J. Song, H. L. Shi, Z. W. Wang, Z.Chen, H. F. Tian, G. F. Chen, J.G.Guo, H. X. Yang and J. Q. Li, *arXiv*, 2011, 1101.2059.

12. T. M. McQueen, Q. Huang, V. Ksenofontov, C. Felser, Q. Xu, H. Zandbergen, Y. S. Hor, J. Allred, A. J. Williams, D. Qu, J. Checkelsky, N. P. Ong and R. J. Cava, *Phys. Rev. B*, 2009, **79**, 014522.
13. V. Petricek, M. Dusek and L. Palatinus, Institute of Physics, Praha, Czech Republic, 2006.
14. L. Palatinus and G. Chapuis, *J. Appl. Cryst.*, 2007, **40**, 786-790.
15. S. Margadonna, Y. Takabayashi, M. T. McDonald, K. Kasperkiewicz, Y. Mizuguchi, Y. Takano, A. N. Fitch, E. Suard and K. Prassides, *Chem. Commun.*, 2008, 5607-5609.
16. F. Q. Huang and J. A. Ibers, *Inorg. Chem.*, 2001, **40**, 2602-2607.
17. M. Zabel and K. J. Range, *Rev. Chimie Min.*, 1980, **17**, 561-568.
18. S. J. Clarke, P. Adamson, S. J. C. Herkelrath, O. J. Rutt, D. R. Parker, M. J. Pitcher and C. F. Smura, *Inorg. Chem.*, 2008, **47**, 8473-8486.
19. L. Haggstrom, H. R. Verma, S. Bjarman, R. Wappling and R. Berger, *J. Solid State Chem.*, 1986, **63**, 401-408.
20. K. Kuroki, H. Usui, S. Onari, R. Arita and H. Aoki, *Phys. Rev. B* 2009, **79**, 224511.
21. Y. Mizuguchi and Y. Takano, *J. Phys. Soc. Jpn.*, 2010, **79**, 102001.
22. M. J. Pitcher, D. R. Parker, P. Adamson, S. J. C. Herkelrath, A. T. Boothroyd, R. M. Ibberson, M. Brunelli and S. J. Clarke, *Chem. Comm.*, 2008, 5918-5920.
23. C. H. Lee, A. Iyo, H. Eisaki, H. Kito, M. Teresa, F. Diaz, T. Ito, K. Kiho, H. Matsuhata, M. Braden and K. Yamada, *J. Phys. Soc. Jpn.*, 2008, **77**, 083704.